# Musical Mix Clarity Predication using Decomposition and Perceptual Masking Thresholds

Andrew Parker [1,*] and Steven Fenton [2]

[1] Centre for Audio and Psychoacoustic Engineering (CAPE), University of Huddersfield, Huddersfield, HD1 3DH, United Kingdom; Andrew.Parker@hud.ac.uk
[2] Centre for Audio and Psychoacoustic Engineering (CAPE), University of Huddersfield, Huddersfield, HD1 3DH, United Kingdom; S.M.Fenton@hud.ac.uk

\* Correspondence: Andrew.Parker@hud.ac.uk

**Featured Application:** The music mix clarity model proposed in this study is expected to form part of a mix clarity meter for use in music production and music information retrieval applications.

**Abstract:** Objective measurement of perceptually motivated music attributes has application in both target-driven mixing and mastering methodologies and music information retrieval. This work proposes a perceptual model of mix clarity which decomposes a mixed input signal into transient, steady-state, and residual components. Masking thresholds are calculated for each component and their relative relationship is used to determine an overall masking score as the model's output. Three variants of the model were tested against subjective mix clarity scores gathered from a controlled listening test. The best performing variant achieved a Spearman's rank correlation of *rho* = 0.8382 (*p*<0.01). Furthermore, the model output was analysed using an independent dataset generated by progressively applying degradation effects to the test stimuli. Analysis of the model suggested a close relationship between the proposed model and the subjective mix clarity scores particularly when masking was measured using linearly spaced analysis bands. Moreover, the presence of noise-like residual signals was shown to have a negative effect on the perceived mix clarity.

## 1. Introduction

Terms such as 'clarity', 'punch', 'warmth' and 'brightness' are semantics often used to describe perceptual features found in musical mixes. These features are often subconsciously combined by a listener when assessing the overall quality of a musical mix. Whilst some of the perceptual features represented by the semantics outlined have an objective counterpart [1–5], clarity does not.

Clarity in the context of this work is related to Pedersen and Zerarov's sound wheel term 'clean' [6], which is defined as: "It is easy to listen into the music, which is clear and distinct. Instruments and vocals are reproduced accurately and distinctly. The opposite of clean: dull, muddy." Other similar definitions can be found in the literature, for example [7–11]. Although none of these definitions are uniform in wording, they have a similar focus on the separability of the component parts of the mix such that each part is distinctly audible. Potential links between the perception of single instrument clarity and brightness, measured using centroid and harmonic centroid, have been suggested [7, 12]. This work evaluates a model-based approach to objective mix clarity prediction. Perceptually motivated metrics have their use in metering and control applications, as well as automatic mixing [10, 13–15]. Additionally, understanding the underlying related signal characteristics of these perceptual features facilitates the proposal of formalised definitions.

Considering the acoustic characteristics of a space, it is possible to objectively determine the clarity and intelligibility that can be achieved. Measures such as C50 (ratio between early and late arriving reflections) and early decay time (EDT) can be combined for this purpose. The direct to reverberant ratio (D/R) has also been linked to acoustic clarity [16]. Furthermore, in a proposal for objective measurement for loudspeaker quality [17], 'clearness' is considered to be a perceptual dimension. This is calculated based on perceived degradation imparted on a signal by the loudspeaker under test in reference to an 'ideal' loudspeaker signal. This approach is similar to established standards to objectively measure music and speech quality, PESQ [18] and PEAQ [19]. These compare encoded signals with their unencoded counterparts to estimate a perceived error signal caused by encoding, which can then be measured using a number of features to determine how disturbing it is. What is common between these measures is that they all compare a signal with an effected version of itself, for example altered by a room or a loudspeaker's response. In the proposed model, there is not a processed version of the signal under test to compare to. Instead, the signal is decomposed, and a comparison made between its desirable and less-desirable attributes is made.

In previous research [20, 21], it has been suggested that masking between signals constituting a multitrack mix could be related to the perception of clarity of the overall mix. Auditory masking is a phenomenon of hearing in which energy present at the ear is not perceived, due to stronger neighbouring energy in frequency or time, known as frequency masking and temporal masking respectively [22]. The model proposed for mix clarity prediction is based on an analysis of the masking relationship between transient, steady-state, and residual components of the signal, utilising the MPEG Psychoacoustic Model II [23, 24]. The model's performance is assessed using a correlation test against subjective mix clarity scores elicited from a controlled listening test. In addition, further analysis of the model response to an independent dataset consisting of musical excerpts with varying degrees of controlled degradation is undertaken.

## 2. Transient, Steady-State and Residual Masking Model

The proposed model has some similarities to a cross-adaptive approach employed to minimise masking between composite signals of a multitrack mix in automatic mixing systems [10, 14]. The automatic mixing system employed in [10] was shown to increase the perceived clarity of the mixes where the amount of masking had been reduced. To achieve this, a cross-adaptive masking metric was used to calculate the level by which a given signal in the multitrack was masked by the sum of all other signals of the multitrack mix. The signals contributing to the given multitrack mix are then processed in such a way to minimise the amount they are masked, thus lowering the amount of masking occurring in the multitrack mix overall. A similar approach was also proposed as a Hierarchical Perceptual Mixing (HPM) system [25], which first determines the most important signals present in the mix as a function of time based on a user parameter, then calculates the perceptual masking threshold of these dominant signals and removes the masked energy of the non-dominant signals present in the mix. It is suggested this approach may improve the clarity of the resulting mix. However, this was not related to any subjective testing.

The MPEG Psychoacoustic Model II's output, employed in the cross-adaptive automatic mixing system [10], is a Signal-to-Mask ratio (SMR). It indicates the ratio of the energy of the input signal to a masking threshold, which is calculated as a function of frequency and time [23, 24]. The masking threshold is calculated by grouping spectral lines into threshold calculation partitions, which represent approximately a third of a critical bandwidth. Energy in these partitions is spread and then weighted based on a tonality measure. The tonality measure is a sliding scale indicating how noise-like or tone-like the input is. More noise-like partitions result in an increased masking threshold as they are more effective maskers. This weighted masking threshold is then compared with the threshold in quiet and the largest of the two is taken as the final masking threshold in each partition. By calculating the masking threshold of an external signal, the SMR reflects the level at which the external signal masks the input signal.

In typical MIR applications this multitrack based approach cannot be used directly as it assumes access to all the signals present which make up the multitrack mix. Additionally, it is unable to calculate masking occurring contained within a single signal, for example if multiple instruments were captured by a single microphone. Whilst there is an emerging field of source separation methods utilising deep learning techniques to 'unmix' mixed signals, such as Spleeter [26]; these systems are unsuited in this case as they are only able to classify a limited number of instruments, resulting in the cross-masking being poorly represented in cases where the given mix is split into only 1 or 2 instrument categories.

*2.1. Model Details*

The proposed model assumes no knowledge of the individual signals making up the multitrack mix. It utilises a novel approach which calculates the level of perceived masking between the separated transient, steady-state, and residual components of the multitrack mix.

When considering a spectrogram representation these components are characterised as follows:

- Transient components: Spectrally broadband and temporally transient bursts of energy which form strong vertical beams on the magnitude spectrogram [27], and unpredictable changes in phase [28].
- Steady-state components: Slowly evolving harmonic partials with predictably evolving phase [28] forming strong horizontal beams on the magnitude spectrogram [27].
- Residual components: Cannot be classified as transient or steady-state. These are noise-like components of the signal, described as the 'texture' of the sound [29], which are generally broadband and stochastic in terms of magnitude and phase.

The TSR model gives a more generalised description of signal components that do not require specific instrument classification. Following the suggested negative impact of noise-like signals on mix clarity [20, 21], greater masking of the residual (R) component would indicate a less audible residual component and therefore a potentially higher perceived mix clarity. Additionally, and perhaps more importantly, less masking of the transient and steady-state (TSS) components would indicate greater audibility of the percussive (rhythmic) and harmonic (pitched) parts of the signal. Transient onsets have also been linked to instrument identification [30], suggesting potentially greater mix clarity perception where TSS components are less masked. Considering the HPM system [25], masked residual energy bears some resemblance to the idea of masked non-dominant signals, whose presence may be unnecessary or even detrimental to the perceived clarity of the mixed signal. Moreover, the importance of TSS and R components can be thought of as a simple hierarchy, where presence of the TSS components take priority over R components. A block diagram of the proposed clarity model is given in Figure 6.

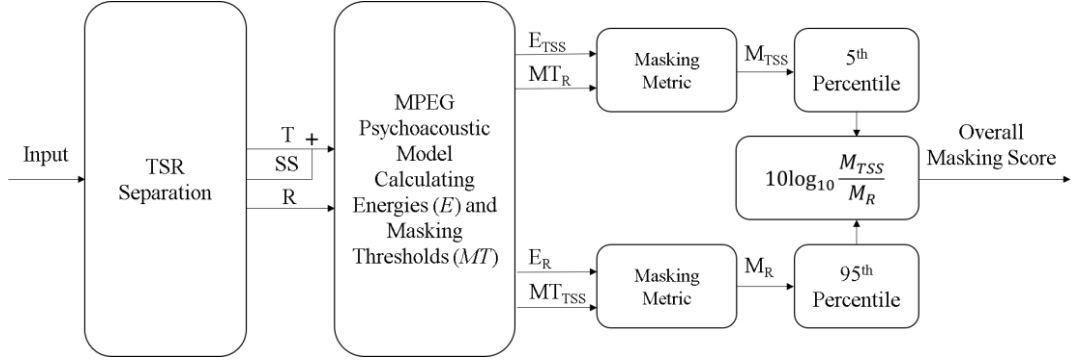

**Figure 6.** Block diagram of the proposed transient, steady-state and residual based masking metric.

The TSR separation system is a median filter based approach [29], chosen for its prior good performance in the perceptual measure of punch [1, 31] and its relative simplicity. In this case, a single layer of filtering on a short time Fourier transform (STFT) spectrogram consisting of a 2048 sample window with 50% overlap, where the separation parameters were a percussive threshold of 1.75, a harmonic threshold of 1, and a median filter had an order of 17. These parameter values were chosen based on previous works, namely the perceptual punch meter [35]. The separated transient and steady-state components can simply be combined by addition to construct the TSS signal without the residual component.

A signal-to-mask ratio (SMR) is calculated by the MPEG Psychoacoustic Model II, measuring the level of masking present as a function of frequency and time [24, 32]. This is the ratio of the energy in each scale-factor band ($E$) to a calculated masking threshold ($MT$). Thus, the SMR is defined as:

$$SMR(sb) = \frac{E(sb)}{MT(sb)} \qquad (1)$$

In the case of the cross-adaptive masking metric [10], the energy and threshold calculation for the scale-factor bands is kept the same, though they define a masker-to-signal ratio (MSR) given in decibels as:

$$MSR(sb) = 10 \log_{10}\left(\frac{MT'(sb)}{E(sb)}\right) \qquad (2)$$

Where $MT'$ is the threshold calculated for the sum of accompanying signals. They assume masking occurs in any band where $MT' > E$, and scale the outcome by a predefined *Tmax* value of 20dB, giving the final masking metric as:

$$Masking\ Metric = \sum_{sb \subset E < MT'} \frac{MSR(sb)}{Tmax} \qquad (3)$$

The model proposed in this paper incorporates the MPEG Psychoacoustic Model II to calculate energy and masking thresholds of the TSS and R components of the separated input signal. Masking metrics are then calculated based on the cross-adaptive metric [10] for these signals in parallel, indicating the amount the R component is masked by TSS component and vice versa. Giving masking metrics $M_{TSS}$ and $M_R$:

$$\begin{aligned} MSR_{TSS}(sb) &= 10 \log_{10}\left(\frac{MT_R(sb)}{E_{TSS}(sb)}\right) \\ MSR_R(sb) &= 10 \log_{10}\left(\frac{MT_{TSS}(sb)}{E_R(sb)}\right) \end{aligned} \qquad (4)$$

$$M_{TSS} = \sum_{sb \subset E_{TSS} < MT_R} \frac{MSR_{TSS}(sb)}{Tmax}$$
$$M_R = \sum_{sb \subset E_R < MT_{TSS}} \frac{MSR_R(sb)}{Tmax} \tag{5}$$

A statistical representation of each metric is taken to represent an overall score for the given analysis window, in this case 10 seconds, as this was the length of the stimuli included in the listening test. These are: the 5th percentile of $M_{TSS}$ giving the points where the TSS component is minimally masked, and the 95th percentile of $M_R$ giving the points where the R component is maximally masked. The overall masking metric is defined as the ratio between the statistical representations of masking expressed in decibels:

$$Overall\ Masking\ Score = 10\log_{10}\left(\frac{5thPercentile(M_{Tss})}{95thPercentile(M_R)}\right) \tag{6}$$

As such, where $M_R$ is low and $M_{TSS}$ is high a large overall score is given and in the opposing case a low overall score is given, thus the score is negatively correlated with mix clarity perception. The layer II (L2PM) and layer III (L3PM) implementations of the MPEG Psychoacoustic Model II present some differences in the calculation of the masking threshold and SMR, though they are both based on the same principles [23]. Both implementations along with a L3PM variant were employed in the model; each was tested and are examined in this work.

It is worth noting, while ideally the separation system incorporated in the model would leave no residual energy in the TSS component and vice versa, this is not the case. In practice, with a purely white noise input, the output of energy to both TSS and R components is approximately even. We would expect in the case of ideal separation that only the R component would contain the signal energy, however, this is not an issue in the present application as the white noise example simply represents the upper-bound of overall masking scores where more structured signals are better separated. This results in a lower overall masking score.

### 3. Test Stimuli

The model outputs were evaluated against subjective scores collected in a controlled listening test. The stimuli used in testing were widely stylistically varied in order to determine features of clarity that applied regardless of instrumentation or arrangement, such that any resulting metrics may be generally applicable across a wide range of music. In addition, an independent dataset was synthesised to investigate the effect of specific forms of signal degradation on the model's output.

*3.1. Subjective Score Elicitation Test*

A controlled listening test was conducted to elicit perceived mix clarity scores across a selection of stylistically different musical stimuli as presented in previous work [20]. 18 listeners took part in the listening test, of which 11 were undergraduate students enrolled on a music technology course, 4 were postgraduate researchers, and 3 were Doctors in the field of Psychoacoustics.

**Table 1.** Table of selected listening test stimuli

| FMA ID | Parent Genre | Sub-Genre |
|---|---|---|
| 144179 | Hip-Hop | Alternative Hip-Hop |
| 134826 | | Hip-Hop |
| 021895 | Rock | Indie-Rock |
| 040903 | | Rock |
| 067357 | Experimental | Audio College |
| 081895 | | Field Recording |
| 053727 | International | Polka |
| 062596 | | Afrobeat |
| 067357 | Pop | Experimental Pop |
| 121592 | | Pop |
| 094414 | Electronic | Ambient Electronic |
| 114556 | | Electronic |
| 122356 | Folk | Singer-Songwriter |
| 146639 | | Psych-Folk |
| 126410 | Instrumental | Soundtrack |
| 137167 | | New-Age |

The test stimuli were taken from the Free Music Archive (FMA) 'FMA small' dataset [33], which consists of 8000 30-second long musical excerpts from 8 different parent genres. The dataset was processed and a random selection method was used to select 16 10-second long 44.1kHz 16 bit wave file stimuli which were monophonic and loudness normalised (-23LU) [5]. This process allowed the selection of 16 stimuli that were widely stylistically varied without the need for any personal selection. Details of the selected stimuli are given in Table 1.

*3.2. Procedure*

Listeners were asked to rate the stimuli for perceived mix clarity. This was a standalone and absolute judgement of mix clarity without reference or comparison to any other piece of music, similar to how mix clarity would be judged were the listener to come across a piece of music on the radio or a music playlist. The custom test interface used was designed in MAX [34] by modifying an existing HULTIGEN [35] interface. Each stimulus was presented individually along with a slider on which the listener was asked to rate the mix clarity they perceived between 0 and 100 in steps of 1. A score of 0 indicated the stimulus was perceived to be unclear (no clarity), and a score of 100 indicated the stimulus was perceived to be clear (highest clarity). Labels were given according to a standard 5-point category rating scale [36]. 6 repeats were performed where the order of the stimuli was randomised each time.

*3.3. Results*

The listeners' results were screened based on the consistency of their repeat scores, as a lack of consistency between repeats showed the listener may not have had a consistent perception of mix clarity or may not have understood the task at hand. Repeat consistency was measured using the interclass correlation coefficient (ICC) between a given listeners repeats. Estimated ICCs and their 95% confidence intervals were calculated using SPSS [37]. The ICC was a mean-rating, absolute agreement, 2-way mixed effects model as recommended in [38]. Ratings from listeners whose repeat scores that had an ICC of below 0.75 were excluded from the final analysis, as it is suggested ratings with an ICC equal to or greater than 0.75 have a 'good' reliability [38]. In this case, the lower bound of the ICC estimates' 95% confidence intervals were taken as the value which needed to exceed the 0.75 threshold, as this ensured with 95% confidence the listeners had 'good' reliability. After post-screening, only ratings from the 15 most consistent listeners remained.

The mean of each listener repeat rating was taken as the listener's subjective clarity score for a given stimulus. The median of all listeners' mean ratings for each stimulus was then taken as its median clarity score (MCS). The median clarity scores, along with their approximated 95% confidence intervals [39] are shown in Figure 1.

The assumption of homogeneity of variances for ANOVA was not met, indicated by both Levene's and Barlett's tests showing $p < 0.05$ for all stimuli. As such, a multiple pairwise Wilcoxon test was performed to identify significantly different MCSs. Holm correction was applied to account for multiple comparisons. The results affirm what is indicated by the confidence intervals shown in Figure 1, in that stimuli whose confidence intervals do not overlap are also shown as being significantly different ($p < 0.05$) in the Pairwise Wilcoxon analysis.

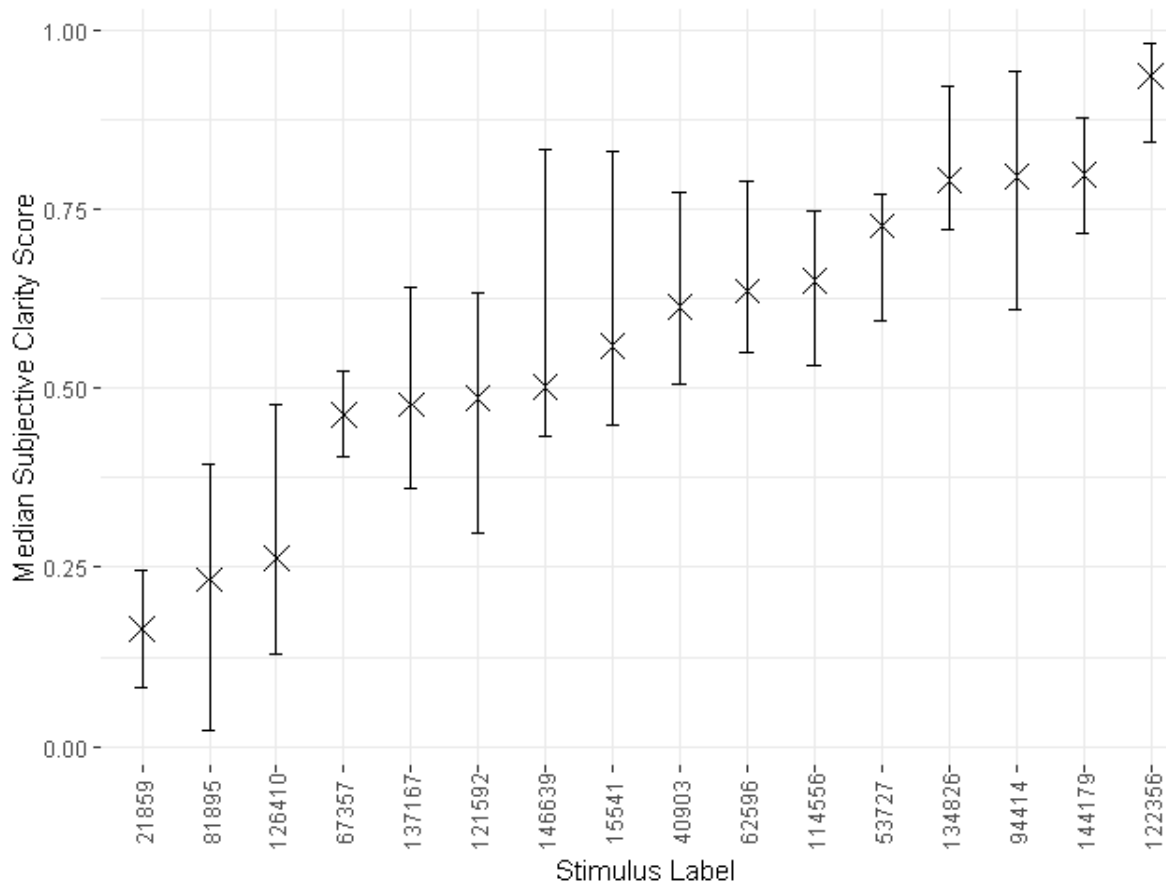

**Figure 1.** Median subjective clarity scores (crosses) and their approximated 95% confidence bounds (bars).

*3.4. Independent Dataset*

An independent dataset was synthesised to investigate the effect of specific signal degradation and how it would impact the predicted clarity score. For this, the stimuli used in the subjective listening test (see Section 3.1) were processed in 3 different ways, with 7 levels of severity, creating 3 test datasets. The processing was intended to gradually decrease the level of perceptual clarity at each level of severity by gradually introducing masking from noise-like signals common in music production. By forming the dataset from the stimuli which had be rated subjectively by listeners, the effect of the degradation on the proposed model could be evaluated relative to the MCSs the stimuli received.

The 3 sets included:

- Set 1 – Addition of broadband pink noise in 6dB steps (-36dBFS, -30dBFS, -24dBFS, -18dBFS, -12dBFS, -6dBFS, 0dBFS). This degradation simulates a gradually raising noise floor or an extremely 'busy' mix where lots of conflicting signals become noise-like.
- Set 2 – Addition of reverberation in 3dB steps of wet mix level (-36dBFS, -30dBFS, -24dBFS, -18dBFS, -12dBFS, -6dBFS, 0dBFS). In this case, the reverberation was meant to simulate a large hall, and thus had a diffuse tail with a decay time of 3 seconds, with decay times more pronounced for low frequency energy rather than high frequency energy. Artificial reverberation is commonly added in music production to enhance the sense of space and listener envelopment. However, it also disperses diffuse energy over time adding to the signal's noise floor and can smear transient energy over time making them less clearly defined.

- Set 3 – Clipping applied at a ceiling calculated as a percentile range of amplitude values about the fiftieth percentile (90%, 80%, 70%, 60%, 50%, 40%, 30%), such that clipping applied is uniform regardless of signal amplitude. Clipping is often used as a creative effect, though also occurs in signals recorded without an appropriate level of headroom. This can cause a reduction in dynamic range, decreasing the energy difference between the signal peaks and the noise floor. In addition, clipping can produce additional harmonic content making the signal more spectrally dense, thus increasing the potential for masking to occur. However, increasing spectral density and brightness of transient and steady-state components may be beneficial to the perception of clarity in some cases, whereby the masking potential of the transients and steady-state components in the signal are increased greater than the masking potential of the residual component.

Each of the test signals generated were loudness normalised to -23LU after the signal degradation had been applied.

4. **Results**

To evaluate the model's performance, the outputs of the clarity model variants were correlated against the MCSs obtained from the controlled listening test detailed in Section 3.3. Both Pearson's correlation coefficient *r* and Spearman's rank correlation coefficient *rho* were calculated as a lack of a bivariate normal distribution or linear relationship between the two variables can cause the Pearson's coefficient to provide an inaccurate measure of association. Pearson's coefficient is still provided, as a non-linear relationship is suggested in cases where *rho* is greater than *r*, and a linear relationship where *r* is greater than *rho*. The Spearman's rank coefficient is non-parametric, so is more robust, and is the focus of the following analysis. Both correlation coefficients for each model variation is given in Table 2 for the sake of comparison.

**Table 2.** Pearson and Spearman rank correlations of tested metrics against MCSs

| Measure | r | rho |
|---|---|---|
| L3PM Clarity Model | -0.6682 | -0.6882 |
| Modified L3PM Clarity model | -0.7868 | -0.8088 |
| L2PM Clarity Model | -0.7884 | -0.8382 |

All coefficients shown are significant where $p < 0.01$.

When correlating the L2PM clarity model masking scores against the MCSs, a strong negative Spearman rank correlation of $rho = -0.8382$, $p < 0.01$ was achieved. This correlation is shown in Figure 2.

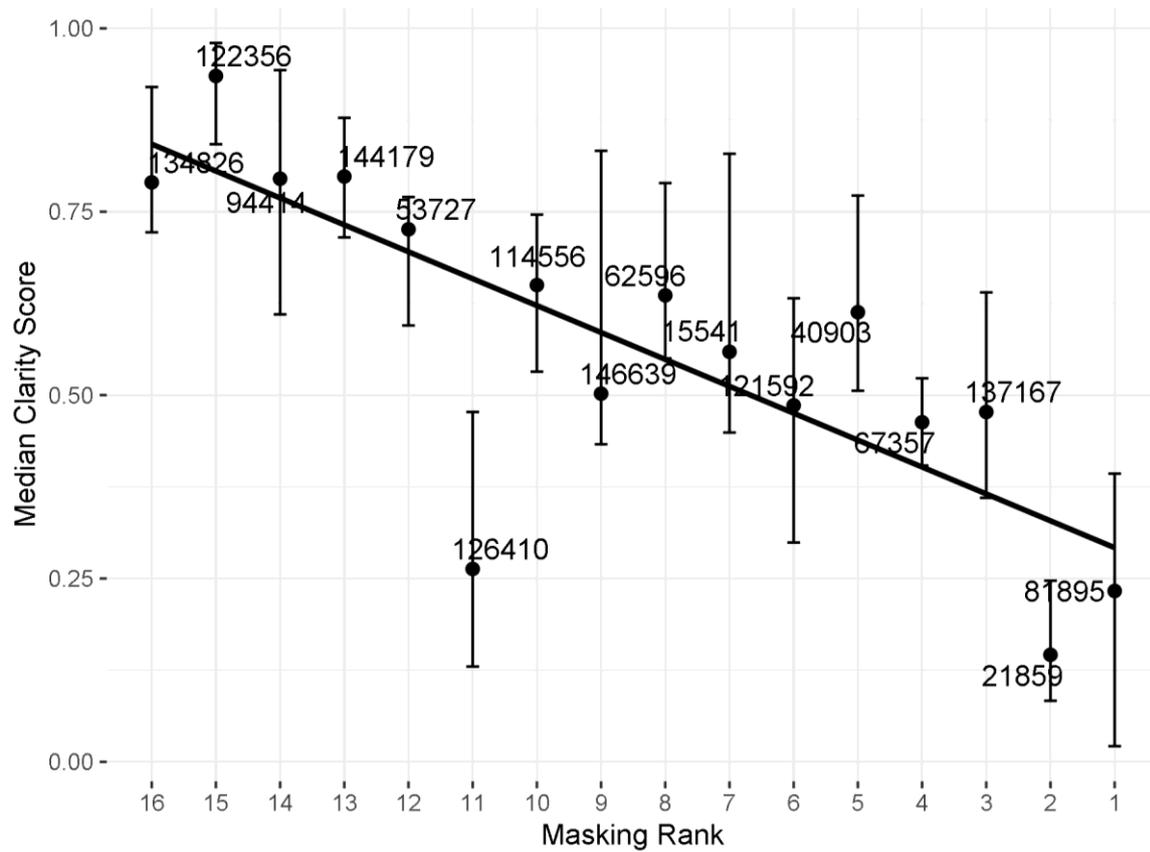

**Figure 2.** Spearman Rank correlation between overall masking score calculated using the layer II implementation of the Psychoacoustic Model II and MCSs collected via controlled listening test.

This was the highest Spearman correlation achieved by any implementation of the system, with most deviations from the line of best fit still within the 95% confidence bounds of the MCSs. The least well ranked stimulus, '126410', was also poorly predicted by a different mix clarity model suggested in previous work [20]. This implementation also achieved $r = -0.7884$, $p < 0.01$, which suggests a strong and slightly non-linear parametric relationship between the model output and the subjective scores.

The model was also implemented using the L3PM, forming the L3PM clarity model. This achieved a significant, but weaker correlation of $rho = -0.6882$, $p < 0.01$, shown in Figure 3. Again, a lesser Person correlation ($r = -0.6882$, $p < 0.01$) was seen, indicating a somewhat non-linear relationship.

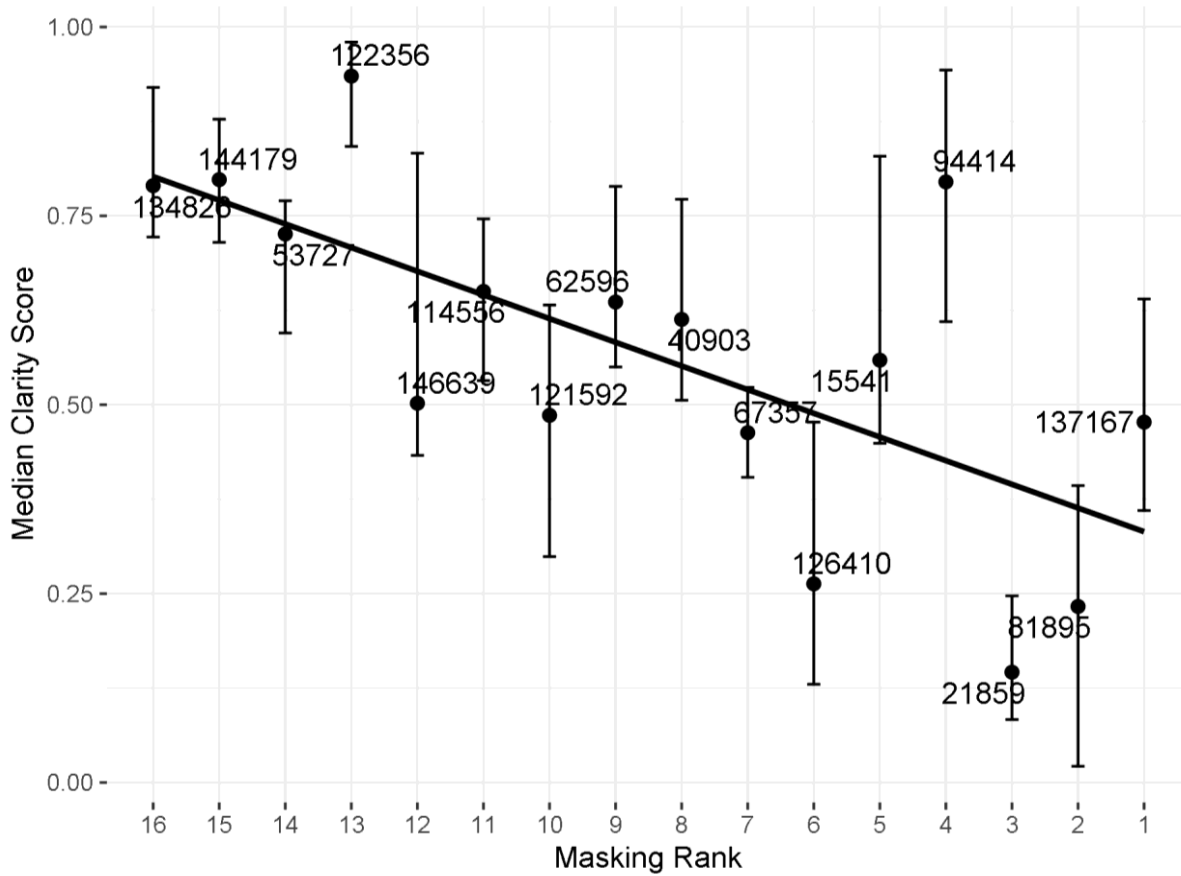

**Figure 3.** Spearman Rank correlation between overall masking score calculated using the layer III implementation of the Psychoacoustic Model II and MCSs collected via controlled listening test.

This correlation shows some heteroscedasticity, where the stimuli given higher masking scores were predicted less accurately than those given low masking scores, though many of the 95% confidence intervals cross the line of best fit. The weaker correlation of this model compared to the L2PM clarity model was somewhat unexpected, as the L3PM is more efficient in coding applications, and the SMR values are calculated in scale-factor bands which approximate critical bands. These are more reflective of perception than the L2PM scale-factor bands which are linearly spaced. The better performance of the L2PM clarity model suggests a greater importance of masking occurring between higher frequency energy, due to the model calculating the overall masking of a given frame in each component as the sum of masking within the scale-factor bands (see Section 2, Equation 5). Other work has suggested an importance of high frequency energy in clarity perception for some single instrument sounds [7, 12]. The L3PM also employs window switching which the L2PM implementation does not [24]; whereby, shorter analysis windows and fewer scale-factor bands are used in determining the energy and masking thresholds of highly transient frames. For the stimuli used in the present testing, the perceptual entropy threshold was not crossed by any frames of any of the stimuli, therefore short analysis frames were not used in any of the masking calculations and thus could not be responsible for the difference in performance between the L2PM and L3PM clarity model variants.

To confirm that the difference in scale-factor bands was largely responsible for the weaker correlation of the L3PM clarity model, a modified version of the L3PM was devised. This calculated the energy and masking thresholds as normal [23, 24], however rather than calculating SMR directly in scale-factor bands which approximate critical bandwidth, the masking thresholds were spread back across the FFT-bins and the SMR was calculated for the linearly distributed bands specified in the L2PM

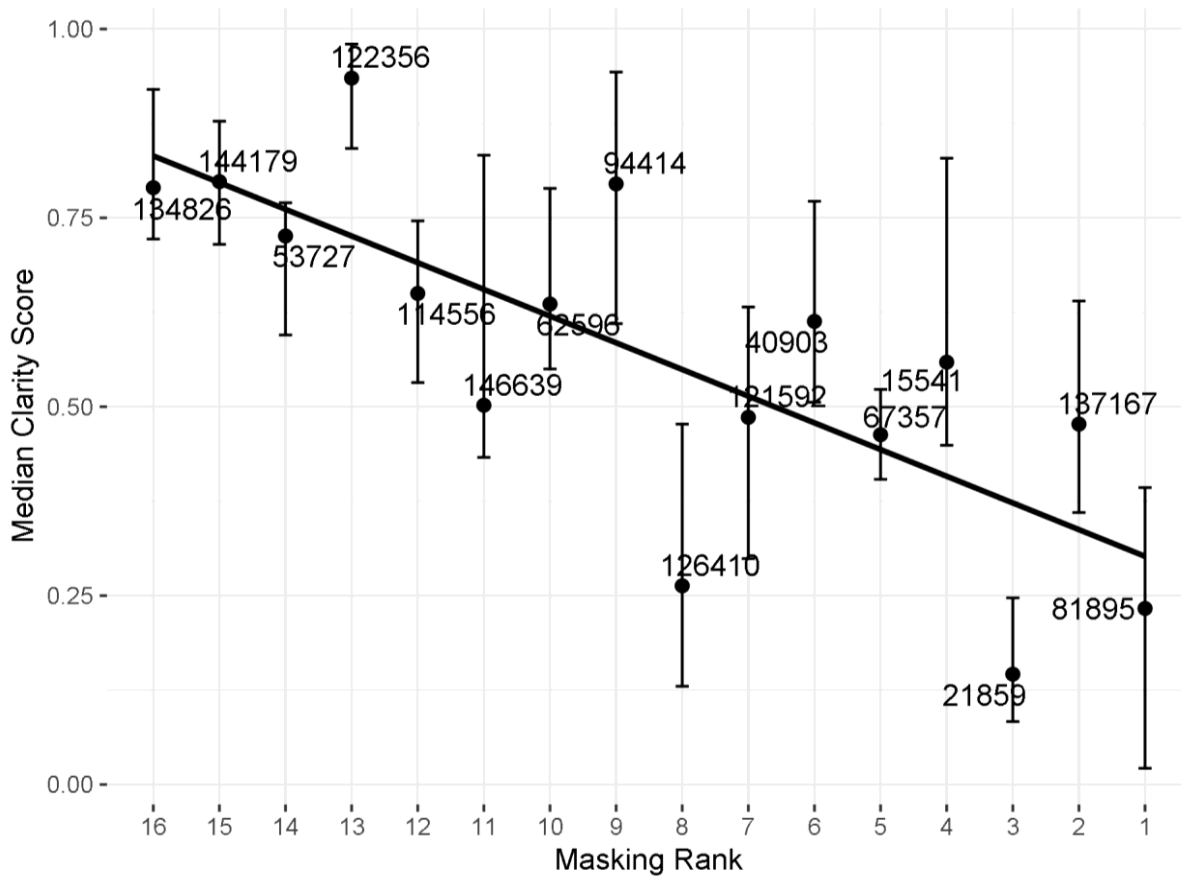

**Figure 4.** Spearman Rank correlation between overall masking score calculated using the modified layer III implementation of the Psychoacoustic Model II and MCSs collected via controlled listening test.

[23]. This saw an increase in performance to a level similar to, but lesser than that of the L2PM clarity model, achieving *rho* = -0.8088, *p* < 0.01 and is shown in Figure 4.

The modified L3PM clarity model improves the position of the L3PM clarity model's most severe outlier, '94414', though a number of outliers remain. The remaining differences between the L2PM and modified L3PM clarity models were due to the difference in how the models calculate the masking threshold. Whist this implementation of the clarity model had a slightly weaker Spearman and Pearson correlation (*r* = -0.7868, *p* < 0.01) than the L2PM clarity model, the Spearman and Pearson correlation coefficient values are similar, suggesting a more linear relationship between the modified L3PM clarity model and the subjective scores.

Further to testing the clarity model variations' correlations to the subjective data, they were also evaluated using the independent dataset (see Section 3.4). Figure 10 shows the scores calculated for audio examples at the various levels of degradation for each of the 4 degradation methods. Box plots show median and interquartile range as well as outliers at each degradation level, the scores are also tied with lines to show the progressive change in model output which occurred as increasing levels of degradation was applied.

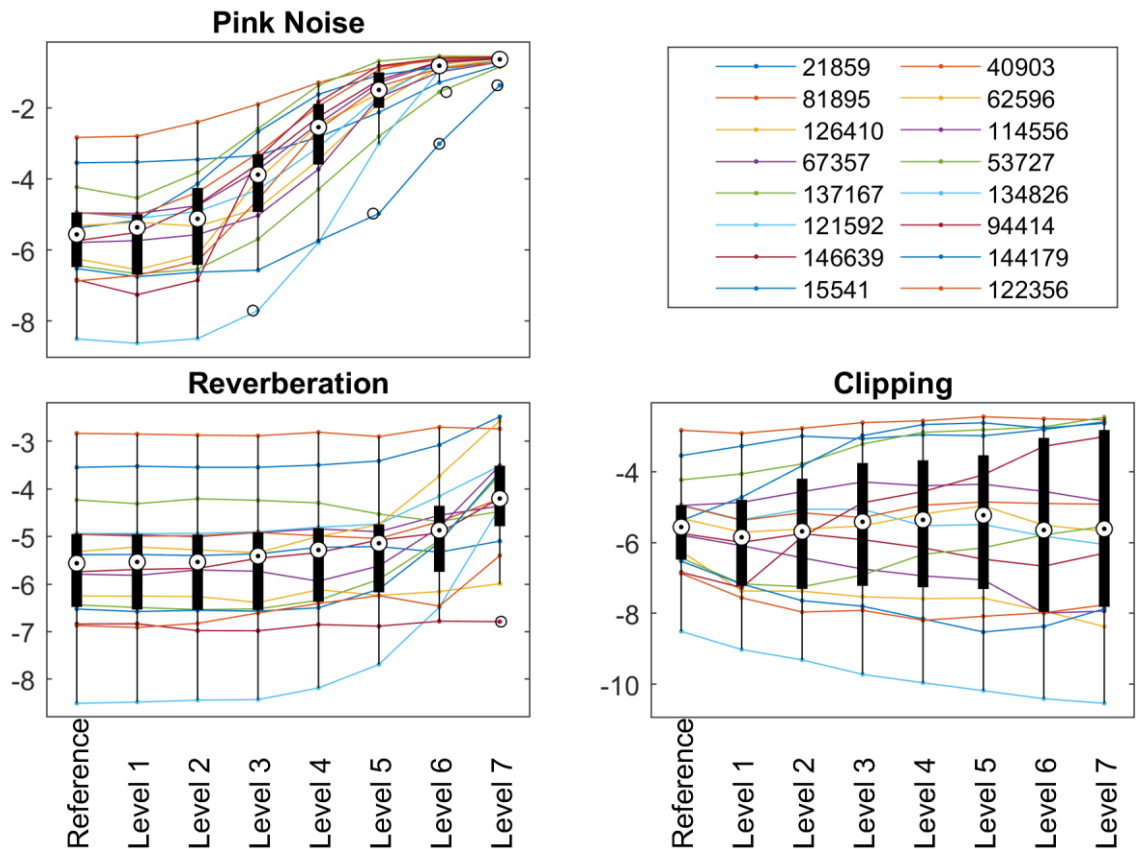

**Figure 5.** L2PM clarity model masking scores of the independent dataset at the 7 levels of degradation for each method (detailed in Section 2.4), including the reference level of the examples without degradation. The legend is listed by MCS ascending column-wise.

Figure 5 shows clarity scores calculated by the L2PM clarity model for the independent dataset. The addition of pink noise sees the addition of broadband residual energy resulting in the masking of transient and steady-state components of the stimuli. As such, the masking scores progressively rise as more noise is added, until they reach a ceiling level where the TSS component is maximally masked. This ceiling causes a convergence of the clarity scores for the tested stimuli, with the tracks scoring a lower MCS in the subjective test tending to see less change when degraded by pink noise. This suggests how noise-like the stimuli are may have had an influence on the MCSs they received by listeners.

The addition of reverberation had a similar though less extreme effect as adding pink noise. Unlike pink noise, the reverberated signal is still derived from and therefore correlated to the original signal. If viewed using a spectrogram, reverberation is somewhat akin to blurring an image where the energy is smeared over time, and to a lesser extent frequency. This smeared energy is largely characterised by the separation system as residual energy, increasing the level to which the TSS component is masked by the R component and vice versa. The addition of reverberation tended to increase the masking score of the stimuli. However, a greater effect is seen for stimuli which received a higher MCS similarly to the addition of pink noise. Moreover, masking scores for stimuli which did not contain strong transient onsets from things such as drum hits, such as '137167','15541', and '94414' also showed less difference than those which did, suggesting the masking effect of reverberation is more severe for transient sounds. This is in line with the greater effect reverberation has on the temporal axis of the spectrogram.

Clipping appears to cause two different responses in the model output whereby some stimuli masking scores increase with degradation level and others decrease. The effect of this is almost symmetrical, keeping the median masking score relative constant throughout the degradation levels whilst the interquartile range increases. Clipping had the effect of reducing $M_R$ (see equation 5), as the dynamic range was decreased and the difference between the R and TSS components was reduced. However, in cases where there was very little residual energy, the $M_{TSS}$ (see equation 5) was also reduced. Clipping can increase the harmonic density of transients which emphasises their spectrally broadband characteristic and thus increases the level of transient energy separated by the separation system. In cases where the present signal is largely steady-state energy, such as a sustained Rhodes piano chord, clipping can emphasise the temporally constant and more slowly evolving nature of steady-state energy and increase the level of steady-state energy separated by the separation system. Moreover, additional harmonics generated from clipping such a signal are spectrally narrowband and temporally constant, and as such are considered to be steady-state by the separation system, further contributing to the level of steady-state energy extracted. Therefore, stimuli that were more noise-like at the reference level, receiving high masking scores, received even higher masking scores as more degradation was applied. Conversely, stimuli that had a low level of residual energy at the reference level received even lower masking scores at higher degradation levels, as the reduction the R component masking was lesser than the reduction of TSS component masking. Whilst the effects of clipping may potentially increase the perception of clarity, given the more complex effect compared to the addition of pink noise or reverberation, the clipping applied at the highest levels of degradation is extreme. The resulting signals are very noise-like, and as such were expected receive high masking scores, which they did not in the case of stimuli that scored middle or low masking scores at the reference level.

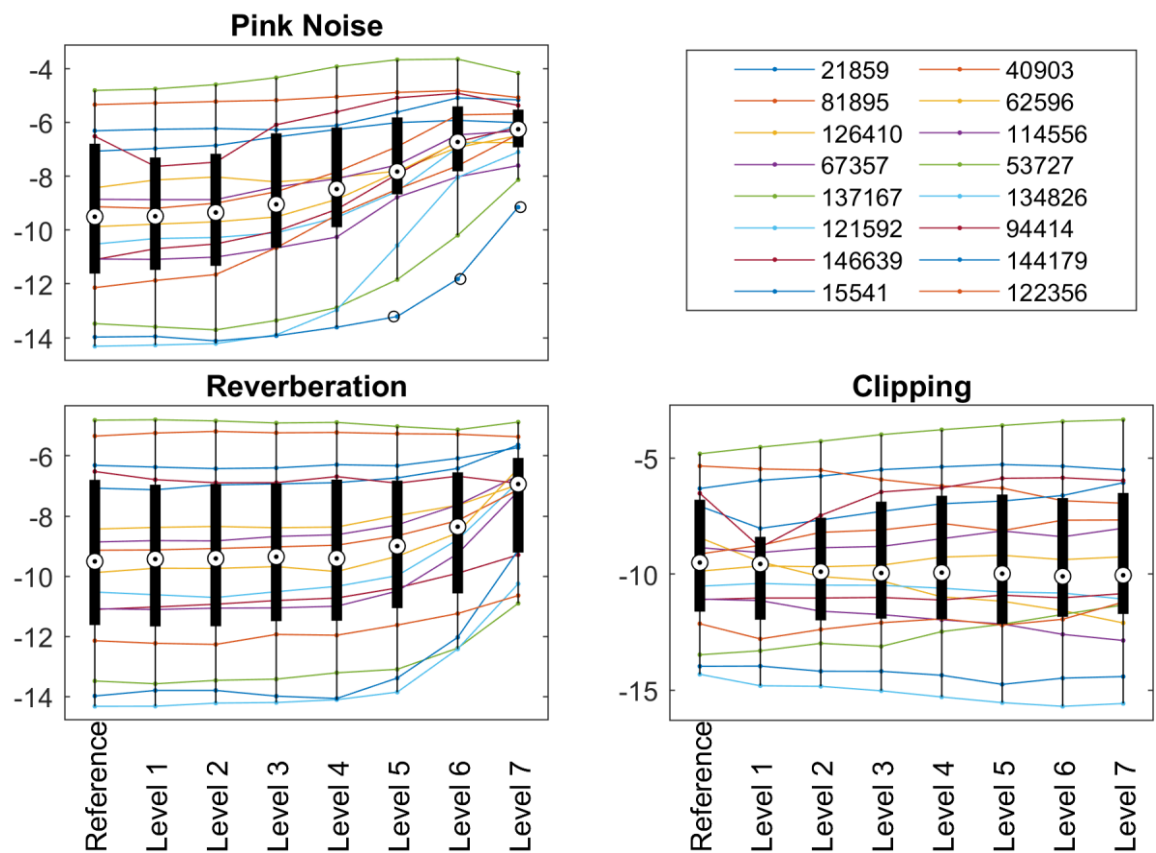

**Figure 6.** L3PM clarity model masking scores of the independent dataset at the 7 levels of degradation for each method (detailed in Section 2.4), including the reference level of the examples without degradation. The legend is listed by MCS ascending column-wise.

The L3PM clarity model responded similarly to the L2PM clarity model for all degradation types. Figure 6 indicates that there was an increase of median masking score and convergence with increasing levels of degradation in the case of additional pink noise and reverberation, and somewhat diverging masking scores with a relatively constant median in the case of increasing levels of clipping. However, this response was muted in comparison to the response seen from the L2PM clarity model. Application of the signal degradation had a greater effect on the separated higher frequency energy than on the separated low frequency energy; this was as a result of the STFT based separation system employed having frequency resolution which increases with bin index. When measured using linearly spaced scale-factor bands, the high frequency bins represent a larger proportion of the scale-factor bands than when measured with logarithmically spaced scale-factor bands. Thus, less change in masking score is seen when degradation is applied in the case of the L3PM clarity model's logarithmically spaced bands, as fewer of these scale-factor bands correspond to the high frequency bins which have the greatest difference in separated energy between degradation levels. While the logarithmic scale-factor bands are more aligned with human perception, in this case, a negative effect on correlation to the subjective scores was seen (see Figures 1 & 2).

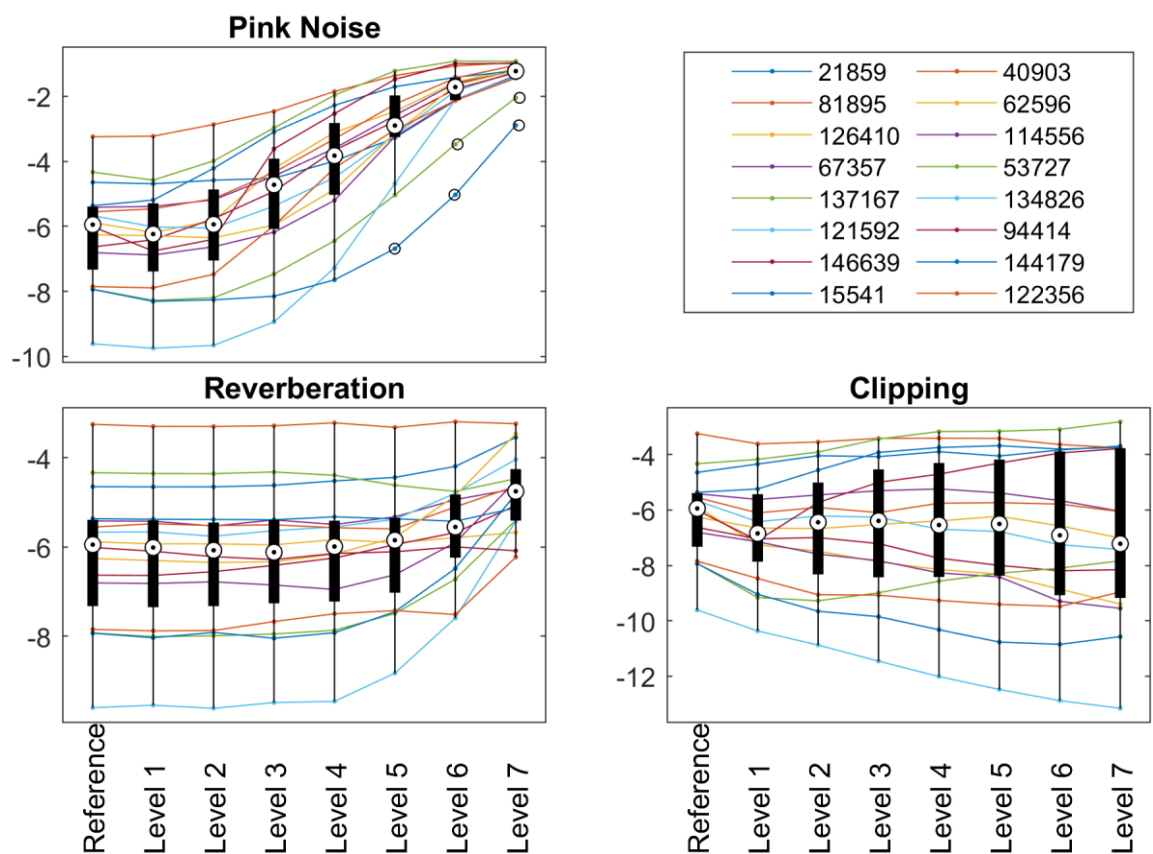

**Figure 7.** Modified L3PM clarity model masking scores of the independent dataset at the 7 levels of degradation for each method (detailed in Section 2.4), including the reference level of the examples without degradation. The legend is listed by MCS ascending column-wise.

Figure 7 shows the clarity model which employed the modified L3PM clarity model, using linearly grouped scale-factor bands like those used in the L2PM clarity model, responded to the degradation similarly to the L2PM clarity model. The similarity of these results, and difference to the L3PM clarity model results (see Figure 6), suggests that the differences between the models' responses could have been caused by the different scale-factor bands employed. The remaining differences then are due to the difference in masking threshold calculation between the L2PM and L3PM clarity models discussed previously [23].

## 5. Discussion

All tested correlations showed a significant and good correlation (<-0.6) to the subjective scores, suggesting a potential link between the underlying concept of transient, steady-state and residual masking and mix clarity perception. However, as the tested subjective dataset contained only 16 stimuli, the model may be overfit, and a larger test containing more participants and stimuli should be performed to validate these results.

The L2PM clarity model showed the strongest *rho* correlation, though was non-linear, with a Spearman's rank correlation test showing a stronger relationship than the Pearson correlation. This implementation of the model is also the least computationally expensive, as the L2PM is less complex than the L3PM. During independent testing, the model responded as expected to the addition of pink noise and reverberation, with the masking score increasing (clarity decreasing) as the degradation levels increased. The model's response to clipping was less uniform than that of the additional pink noise or reverberation, showing a more complex effect on the relationship of the R and TSS components of the signal. While the response is understandable in terms of how the model operates, it is not expected to be congruent with perception. Low perceived clarity scores, corresponding to high masking scores, would be uniformly expected for all stimuli at the highest levels of clipping degradation, showing a potential shortcoming of the model. However, given the strong correlation to the subjective scores, this shortcoming did not seem to have a greatly detrimental effect for the tested stimuli in this case. Though, improving the response to this kind of degradation would improve the robustness of the model in extreme cases.

While the L3PM has a more complex calculation of masking threshold and provides more efficient encoding in the context of MPEG compression compared to the L2PM, the L3PM clarity model unexpectedly had the weakest correlation to the subjective scores. It is suggested this was largely due to L3PM's logarithmically spaced scale-factor bands not providing emphasis on masking occurring in high frequency bands like in the L2PM and modified L3PM variations of the clarity model. This could indicate a greater importance of masking occurring between high frequency energy to the perception of clarity. In terms of independent testing, the L3PM clarity model had a similar but less extreme response to all three degradation types than the L2PM clarity model. Additionally, the L3PM is capable of calculating masking thresholds using shorter windows when transients occur, providing a higher time resolution and reducing pre-ringing artifacts in MPEG coding. Similarly, this window switching could be used in the clarity model to provide greater temporal resolution for masking occurring relating to transient passages in the signal under test. In the present testing only long windows were used, as none of the stimuli under test had transient content capable of triggering the short frame calculation. If a more appropriate onset detection method was used, the application of shorter windows may improve the clarity model's performance.

To confirm the difference between the L2PM and L3PM clarity model variations' performance was largely due to the scale-factor band grouping, a modified version of the L3PM was devised to employ linearly grouped scale-factor bands like the L2PM. The performance of the modified L3PM clarity model response was similar to the L2PM clarity model in both correlation with the subjective scores, and to the independent dataset, affirming the difference in scale-factor bands was largely responsible for the difference in performance between the L2PM and L3PM clarity models. While the *rho* and *r* coefficients showed a slightly weaker correlation than that of the L2PM clarity model, the coefficients were more similar in value, which indicates a somewhat more linear relationship to the subjective data. Additionally, being based on the L3PM clarity model, this model may also benefit from improving the onset detection system used for window switching.

## 6. Conclusion

A new model for prediction of mix clarity has been proposed, based on the masking relationship between residual, transient and steady-state components of a musical signal. The model consisted of a median filter-based separation system, which feeds the MPEG Psychoacoustic Model II used to calculate signal-to-mask ratios of the component parts which are then compared. Both layer 2 and layer 3 implementations of the MPEG Psychoacoustic Model II were tested, along with a modified version of the layer 3 implementation, forming L2PM, L3PM, and modified L3PM variants, respectively.

Each variation was evaluated through both Pearson and Spearman's rank correlation to subjective scores gathered in a controlled listening test, and, though their response to an independent dataset of stimuli degraded though the addition of pink noise, reverberation, and clipping. The L2PM clarity model showed the strongest correlation to the subjective scores, followed by the modified L3PM, with the L3PM clarity model showing the weakest relationship. Although the L3PM is most efficient in coding applications, the stronger correlation achieved by the modified L3PM clarity model showed that the linearly grouped scale-factor bands of the L2PM were advantageous to performance in this case. All variations of the model responded similarly to the degradation introduced in the independent dataset; the L3PM clarity model's response was less extreme to that of the modified L3PM and L2PM clarity models, whose responses were very similar. Addition of pink noise and reverberation caused an increase in masking score, reflecting a decrease in clarity. Clipping caused a somewhat more complex response, where stimuli which were noise-like and received high masking scores at their reference level gained higher masking scores when clipped, and stimuli which had a low level of residual energy and a low masking score at their reference level received even lower masking scores when clipped.

Further work is ongoing to validate the proposed model's performance against a larger subjective data set.